\documentclass{article}
\usepackage{epsfig}
\input epsf.sty
\begin{document}
\title{The saga of the Ising susceptibility}
\author{B.M. McCoy$^1$,  M. Assis$^1$,  S. Boukraa$^2$,
  S. Hassani$^3$, J-M. Maillard$^4$,\\
 W.P. Orrick$^5$, and  N. Zenine$^3$  }
\maketitle
1. CN Yang Institute for Theoretical Physics, State
  University of New York, Stony Brook, NY 11794, USA\\
2. LPTHIRM and D{\'e}partment d'Aeronautique, Universit{\'e} de
Blida. Algeria\\
3. Centre de Recherche Nucl{\'e}aire d'Alger, 2 Bd. Frantz Fanon, BP 399,
16000 Alger, Algeria\\
4. LPTMC, Universit{\'e} de Paris 6, Tour 24, 4{\'e}me {\'e}tage, case
121, 4 Place Jussieu, 75252 Paris Cedex 05, France\\
5. Dept. of Math. Indiana University, Bloomington, Indiana, 47405 USA

\begin{abstract}
We review developments made since 1959 in the search for a closed form
for the susceptibility of the Ising model. The  expressions for the form
factors in terms of the nome $q$ and the modulus $k$ are compared and
contrasted. The $\lambda$ generalized correlations 
$C(M,N;\lambda)$ are defined and explicitly computed in terms of theta
functions  for $M=N=0,1.$
\end{abstract}

\section{Introduction}

There are three important thermodynamic properties of any magnetic
system in zero magnetic field: the partition function from which free
energy and the specific
heat are obtained; the magnetization; and the magnetic susceptibility.
For the two dimensional Ising model in zero field defined
by
\begin{equation}
{\mathcal E}_0\,=\, \,-\sum_{j,k}\{E^v
\sigma_{j,k}\sigma_{j+1,k}+E^h\sigma_{j,k}\sigma_{j,k+1}\}
\end{equation}
with $\sigma_{j,k}=\pm 1$
the free energy was first computed by Onsager~\cite{ons1} in 1944 
and the spontaneous
magnetization was announced by Onsager in 1948~\cite{ons2} and proven
by Yang~\cite{yang} in 1952. To this day a closed form for the
magnetic susceptibility has never been found. We will here trace
the saga of the quest for this susceptibility.

If we could solve the Ising model in the presence of a magnetic field
$H$ which interacts with the total spin of the system as
\begin{equation}
{\mathcal E}=\, {\mathcal E}_0-H\sum_{j,k}\sigma_{j,k}
\label{isingdef}
\end{equation}
then the magnetic susceptibility would be computed as
\begin{equation}
 \chi(H)\,=\,\,\frac{\partial M(H)}{\partial H}
\label{chih}
\end{equation}
where the magnetization is 
\begin{equation}
M(H)\,=\,\, \frac{1}{Z(H)}\sum_{\sigma_{j,k}=\pm 1}\,\sigma_{0,0}\,e^{-{\mathcal E}/k_BT}
\label{mh}
\end{equation}
with the partition function defined by
 \begin{equation}
Z(H)\,=\,\,\sum_{\sigma_{j,k}=\pm 1}e^{-{\mathcal E}/k_BT}
\label{zh}
\end{equation}
However, because the Ising model has only been solved for $H=0$ we are
forced to restrict our attention to $\chi(0)$ which from
(\ref{isingdef})-(\ref{zh}) is given in terms of the two point
correlation functions as
\begin{equation}
k_BT \cdot \chi(0)\,=\,\,\sum_{M,N}\{\langle\sigma_{0,0}\,\sigma_{M,N}\rangle-M(0)^2\}
\label{chidef}
\end{equation}
where $M(0)$ is the spontaneous magnetization of the system which is
zero for $T>T_c$ and for $T< T_c$
\begin{equation}
M(0)\,=\,\, (1-k^2)^{1/8}
\end{equation}
where
\begin{equation}
k\,=\,\, (\sinh 2K^v \sinh 2K^h)^{-1}
\label{km}
\end{equation}
with $K^{v,h}=\, E^{v,h}/k_BT$
and $T_c$ is defined by
\begin{equation}
k=\,1.
\end{equation}

The first exact result for the susceptibility was given in 1959
 by Fisher~\cite{fisher} who used results of Kaufmann 
and Onsager~\cite{ko} to argue that
as $T\rightarrow T_c$ the susceptibility diverges as $|T-T_c|^{-7/4}$.
The saga may be said to begin with  the concluding remark of this
paper:

{\it In conclusion we note that the relatively simple results (1) and
  (3) suggest strongly that there is a closed expression for the susceptibility
in terms of elliptic integrals. It is to be hoped that such a formula
will be discovered,$\cdots$.}

One year later Syozi and Naya\cite{sn}, 
on the basis of short series expansions,
proposed such a formula for $T>T_c$ which does not involve elliptic integrals 
\begin{equation}
k_BT \cdot \chi(0)\,=\,\,\frac{(1-\sinh^2 2K^v\sinh^2 2K^h)^{1/4}}
{\cosh 2K^v \cosh 2K^h-\sinh 2K^v -\sinh 2K^h} 
\label{syozi}
\end{equation}
However, when higher order terms
were computed this conjecture was shown not to be exact.

To this day the  ``closed expression'' for the susceptibility
 hoped for in~\cite{fisher} has not been found.
 
\section{Form factor expansion and the $\lambda$ extension}

To proceed further a systematic understanding of the two point correlation
function is required. For short distances the correlations are well
represented by determinants~\cite{ko,mpw} whose size grows 
with the separation of the spins. However, in order to execute the sum
over all separations required by (\ref{chidef}) an alternative form of
the correlations which is efficient for large distances is
needed. The study of this alternative form was initiated in 1966 
by Wu~\cite{wu} who discovered that for the row
 correlation $\langle \sigma_{0,0}\sigma_{0,N}\rangle$ that when \\
 $N \cdot |T-T_c| \gg \, 1$ for $\,T<T_c$
\begin{equation}
\langle\sigma_{0,0}\sigma_{0,N}\rangle\,
=\,\, (1-t)^{1/4}\cdot \{1+f^{(2)}_{0,N}+\cdots\}
\label{helpm}
\end{equation}
with $t=k^2$ with $k$ given by (\ref{km}). 
For $T>T_c$ we define $k=\, \sinh 2K^v \sinh 2K^h$ and find
\begin{equation}
\langle\sigma_{0,0}\sigma_{0,N}\rangle\, 
=\, (1-t)^{1/4}\cdot \{f^{(1)}_{0,N}\, +\, \cdots\}
\label{helpp}
\end{equation}
with $t=k^{2}$. In (\ref{helpm}) and (\ref{helpp})   
$f^{(n)}_{0,N}$ is an $n$ fold integral 
which exponentially decays for large $N$.
The results (\ref{helpm}) and (\ref{helpp}) are 
the leading terms in what has become known as the form
factor representation of the correlations
 which in general for $T<T_c$ is
\begin{equation}
\langle\sigma_{0,0}\sigma_{M,N}\rangle
\,=\,\, (1-t)^{1/4}\cdot \{1+\sum_{n=1}^{\infty}f^{(2n)}_{M,N}\}
\label{ffm}
\end{equation}
and for $T>T_c$ 
\begin{equation}
\langle\sigma_{0,0}\sigma_{M,N}\rangle
\,=\,\,(1-t)^{1/4} \cdot \sum_{n=0}^{\infty}f^{(2n+1)}_{M,N}
\label{ffp}
\end{equation}
where $f^{(n)}_{M,N}$ is an $n$ dimensional integral. For general
$M,~N$ these $f^{(n)}_{M,N}$ were computed in 1976 by Wu, McCoy, Tracy
and Barouch~\cite{wmtb} and related forms are given in
\cite{pt,yam}. However, for the diagonal
correlations an alternative and simpler form is available which was
announced in~\cite{mccoy1} and proven in~\cite{mccoy2}. For the
diagonal form factor for $T<T_c$
\begin{eqnarray}
&&f^{(2n)}_{N,N}(t) =\, {t^{n(N+n)}\over (n!)^2\pi^{2n}}
\int_0^1\prod_{k=1}^{2n}dx_kx_k^N\prod_{j=1}^n\left(\frac{
(1-tx_{2j})(x_{2j}^{-1}-1)}{(1-tx_{2j-1})
(x_{2j-1}^{-1}-1)}
\right)^{1/2}\nonumber\\
&&\prod_{1\leq j \leq n}\prod_{1\leq k \leq n}
\left({1\over 1-tx_{2k-1}x_{2j}}\right)^2
\prod_{1\leq j<k \leq
  n}(x_{2j-1}-x_{2k-1})^2(x_{2j}-x_{2k})^2\nonumber\\
\label{dffm}
\end{eqnarray}
for $T>T_c$
\begin{eqnarray}
&&f^{(2n+1)}_{N,N}(t)\,=\nonumber\\
&&{t^{(n+1/2)N+n(n+1)}\over
      n!(n+1)!\pi^{2n+1}}\int_0^1\prod_{k=1}^{2n+1}dx_kx_k^N
\prod_{j=1}^{n+1}x_{2j-1}^{-1}[(1-tx_{2j-1})(x^{-1}_{2j-1}-1)]^{-1/2}
\nonumber\\
&&\prod_{j=1}^{n}x_{2j}[(1-tx_{2j})
(x^{-1}_{2j}-1)]^{1/2}\prod_{1\leq j \leq n+1}\prod_{1\leq k \leq n}
\left({1\over 1-tx_{2j-1}x_{2k}}\right)^2\nonumber\\
&&\prod_{1\leq j <k\leq n+1}(x_{2j-1}-x_{2k-1})^2
\prod_{1\leq j <k \leq n}(x_{2j}-x_{2k})^2
\label{dffp}
\end{eqnarray}
In particular 
\begin{equation}
f^{(1)}_{N,N}(t)\,=\,\,t^{N/2}\cdot \frac{\Gamma(N+1/2)}{\pi^{1/2}N!} \cdot 
F\left(\frac{1}{2},N+\frac{1}{2};N+1;t\right)
\label{f1nn}
\end{equation}
where $F(a,b;c;t)$ is the hypergeometric function.

It is often useful and instructive to extend the form factor
expansions (\ref{ffm}) and (\ref{ffp}) by weighting $f^{(n)}_{M,N}$ 
by $\lambda^n$ and thus we define ``$\lambda$ generalized
correlations''
\begin{equation}
C_{-}(M,N;\lambda)
\,=\,\,(1-t)^{1/4} \cdot \{1\,+\sum_{n=1}^{\infty}\lambda^{2n}f^{(2n)}_{M,N}\}
\label{lffm}
\end{equation}
and for $T>T_c$
\begin{equation}
C_{+}(M,N;\lambda)
\,=\,\,(1-t)^{1/4}\cdot \sum_{n=0}^{\infty}\lambda^{2n+1}f^{(2n+1)}_{M,N}
\label{lffp}
\end{equation}
This $\lambda$ extension was first introduced in 1977 by McCoy, Tracy
and Wu~\cite{mtw} in the context of the scaling limit.

\section{Leading divergence as $T\rightarrow T_c$}

These form factor expansions may now be used in (\ref{chidef}) where
the sums over $M,N$ are easily executed under the integral signs to
produce a corresponding expansion of the susceptibility~\cite{wmtb} 
which we write  for $T<T_c$ as
\begin{equation}
k_BT \cdot \chi(0)\,=\,\,\,(1-t)^{1/4} \cdot \sum_{n=1}^{\infty}{\hat \chi}^{(2n)}
\label{chihatm}
\end{equation}
and for $T>T_c$
\begin{equation}
k_BT \cdot \chi(0)\,=\,\,\,(1-t)^{1/4}\cdot \sum_{n=0}^{\infty}\, {\hat \chi}^{(2n+1)}
\label{chihatp}
\end{equation}
where
\begin{equation}
{\tilde \chi}^{(n)}
=\, \sum_{M=-\infty}^{\infty}\sum_{N=-\infty}^{\infty} \, f^{(n)}_{M,N}.
\end{equation}
For $n=1,2$ the ${\hat \chi}^{(n)}$ are explicitly evaluated
\cite{wmtb}. For the isotropic 
lattice we have
\begin{eqnarray}
&&{\hat \chi}^{(1)}\,=\,\,\frac{1}{(1-k^{1/2})^2}\label{chi1}\\
&&{\hat\chi}^{(2)}\,
=\,\,\frac{(1+k^2)\cdot E\, -(1-k^2)\cdot K}{3\pi\,(1-k)(1-k^2)}\label{chi2}
\end{eqnarray}
where $K$ and $E$ are the complete elliptic integrals of the first and
second kind.
The form (\ref{syozi}) of Syozi and Naya \cite{sn} with $K^v=K^h$ 
is seen to be the first term in (\ref{chihatp}).  
It is quite clear that unlike the 1959 argument of
~\cite{fisher} the behavior of the susceptibility as $T\rightarrow
T_c$ will be different depending on whether $T$ approaches $T_c$ from
above or below. This dramatic difference was first seen in 1973
 in~\cite{bmw} where it is shown that for  $T\rightarrow T_c\pm$ 
\begin{equation}
k_B \cdot \chi(0)\, \sim\,\,\,  C_{0\pm} \cdot |T-T_c|^{-7/4}
\end{equation}
where
\begin{equation}
C_{0\pm}
\, = \, \, 2^{-1/2} \cdot \coth 2K^v_c\coth 2K^h_c \cdot
 [K^v_c\coth 2K^v_c \, +K^h_c \coth 2K^h_c]^{-7/4} \cdot I_{\pm} \nonumber 
\end{equation}
and $I_{\pm}$ have been numerically evaluated to 52 digits in~\cite{ongp}:
\begin{eqnarray}
&&I_+=1.000815260440212647119476363047210236937534925597789\, \cdots \nonumber \\
&&I_-= {{1} \over {12\pi}} \cdot 1.000960328725262189480934955172097320572505951770117 \, \cdots 
\nonumber
\end{eqnarray}

\section{The singularities of Nickel}

The next advance in the understanding of the analytic structure of the
susceptibility came in 1996 when Guttmann and Enting~\cite{ge},
using resummed high temperature series expansions of the 
anisotropic Ising model,
argued that the susceptibility cannot satisfy a finite order
differential equation and raised the question of the occurrence of a
natural boundary. This natural boundary argument was made very concrete
for the isotropic case in 1999 and 2000 by Nickel~\cite{nickel1}
 and~\cite{nickel2} who analyzed the singularities of the
$n$ fold integrals ${\tilde \chi}^{(n)}$. These integrals, of course,
have singularities at $T=\, T_c$ where the individual correlation
functions $\langle \sigma_{0,0}\sigma_{M,N}\rangle$ have
singularities. However, Nickel made the remarkable discovery that the
integrals, for ${\hat \chi}^{(n)}$, contain many more singularities. In
particular he found that, for the isotropic lattice, ${\hat \chi}^{(n)}$
has singularities in the complex temperature 
variable $s=\,\sinh 2E/k_BT$ at
\begin{equation}
s=\,\,s_{j,k}=\,\, e^{i\theta_{j,k}}
\end{equation}
where
\begin{equation}
2\cos(\theta_{j,k})\,=\,\,\cos(2\pi k/n)+\cos(2\pi j/n)
\end{equation} 
For $n$ odd $(T>T_c)$ the behavior of ${\hat \chi}^{(n)}$ near the
singularity is
\begin{equation}
 {\hat \chi}^{(2n+1)}\,\sim \,\,\,\epsilon^{2n(n+1)-1} \cdot \ln \epsilon
\end{equation}
with 
\begin{equation}
\epsilon\,=\,\, 1\,-s/s_{j,k}
\end{equation}
and for even $n$ $(T<T_c)$ 
\begin{equation}
 {\hat \chi}^{(2n)}\,\sim\,\,\, \epsilon^{2n^2-3/2}
\end{equation}

The discovery of these singularities demonstrates that the magnetic
susceptibility is a far more complicated object than either the free
energy or the spontaneous magnetization and that the hope expressed
in~\cite{fisher} of a closed form in terms of a few elliptic integrals
is far too simple. 

\section{The theta function expressions of Orrick, Nickel, Guttmann
  and Perk}

In the following year a major advance was made by Orrick,
Nickel, Guttmann and Perk ~\cite{ongp} who studied both the form factors 
and the susceptibility by means of generating on the computer series
of over 300 terms. From these series they then made several remarkable
conjectures for the form factors. 

To present these conjectures we define theta functions as
\begin{eqnarray}
&&\theta_1(u,q)\,=\,\,2\,\sum_{n=0}^{\infty} \,(-1)^n  \, q^{(n+1/2)^2} \cdot \sin[(2n+1)u]\\
&&\theta_2(u,q)\,=\,\,2\,\sum_{n=0}^{\infty} \, q^{(n+1/2)^2} \cdot  \cos[(2n+1)u]\\
&&\theta_3(u,q)\,=\,\,1\,+2\sum_{n=1}^{\infty} \, q^{n^2} \cdot  \cos 2nu\\
&&\theta_4(u,q)\,=\,\,1\,+2\sum_{n=1}^{\infty} \, (-1)^n \, q^{n^2} \cdot  \cos 2nu
\end{eqnarray}
and for $u=0$ we use the short hand
\begin{eqnarray}
\theta_2=\, \theta_2(0,q), \quad \quad  \theta_3 =\,\theta_3(0,q), 
\quad \quad   \theta_4=\,\theta_4(0,q)
\end{eqnarray}

The quantity $q$ is the nome of the elliptic functions and is related
to the modulus $k$ by the relation
\begin{equation}
k\, =\,\, 4\,q^{1/2}\cdot \prod_{n=1}^{\infty}
\left[\frac{1+q^{2n}}{1+q^{2n-1}}\right]^4
\end{equation}

In terms of these theta functions, conjectures for form factors are
given in sec. 5.2 of~\cite{ongp}  by defining an operator
$\Phi_0$ which converts a power series in $z$ to a power
series in $q$ as
\begin{equation}
\Phi_0 \Bigl(  \sum_{n=0}^{\infty} c_n \cdot z^n\Bigr)
 \,\, =\,\,\,\sum_{n=0}^{\infty}c_n \cdot q^{n^2/4}
\label{phidef}
\end{equation}

Conjectures are then given for
$f^{(n)}_{0,0},~f^{(n)}_{1,1},f^{(n)}_{1,0},
~f^{(n)}_{2,0}$ and $f^{(n)}_{2,1}$. 
In particular we note 
\begin{equation}
2^{-n} \cdot (1-k^2)^{1/4}\cdot f^{(n)}_{0,0}
\,\,=\,\,\,\,\frac{(1,k^{-1/2})}{\theta_3}
\cdot \Phi_0\Bigl( \frac{z^n \cdot (1-z^2)}{(1+z^2)^{n+1}}\Bigr)
\label{f00}
\end{equation}
and
\begin{equation}
2^{-n} \cdot (1-k^2)^{1/4}\cdot f^{(n)}_{1,1}\,=\,\,\,
\frac{2\,\, (n+1)(1,k^{-1/2})}{\theta_2\theta_3^2} \cdot 
\Phi_0 \Bigl( \frac{z^{n+1}\cdot (1-z^2)}{(1+z^2)^{n+2}}\Bigr) 
\label{f11}
\end{equation}
where 
\begin{eqnarray}
(1,k^{-1/2})=&&1 \quad \quad {\rm for} \quad \quad T < T_c~~(n~{\rm even})
\nonumber\\
&&k^{-1/2}\quad\quad  {\rm for} \quad \quad T > T_c~~(n~{\rm odd}).
\end{eqnarray}

\section{Linear differential equations}

A second approach to the form factors and  susceptibility  was initiated
in 2004 in ~\cite{maillard1} and subsequently  greatly developed
in ~\cite{maillard2,maillard3,maillard4,maillard5,maillard7,maillard6}. 
These studies are
similar to ~\cite{ongp} in that they expand the form factors and
susceptibility in long series. However, instead of the  nome $q$ 
the expansion is in the (modular) variable $t$. The goal of these
studies is to characterize the $n$ particle contributions
 ${\hat \chi}^{(n)}(t)$ to the susceptibility in terms of finding a Fuchsian
linear ordinary differential equation satisfied by ${\hat \chi}^{(n)}(t)$.
Such a linear differential equation always exists 
for an  $n$-fold integral with
an algebraic integrand in some well-suited choice of integration
variables and in the parameter $t$. However the
order and the degree of the equation rapidly become large for
increasing $n$ and it may take series of many thousands of terms to
find the differential equation. Such a study can only be done by computer.

There are several features of these differential equations to be
noted. In particular the operator which annihilates
 ${\hat \chi}^{(n)}$ factorizes  and furthermore the operator has a direct
sum decomposition such that ${\hat \chi}^{(n-2j)}$ for $j=1,\, \cdots\,  [n/2]$
are ``contained'' in ${\hat \chi}^{(n)}$.
 
\section{Diagonal form factors}

With the  observation of factorization, direct sum decomposition 
and Nickel singularities of the $n$ particle contributions 
to the bulk susceptibility
${\hat \chi}^{(n)}$, it has become clear that
the susceptibility is far more complicated than what was envisaged by
Fisher~\cite{fisher}  in 1959. Because of this complexity the question
was asked if there could be a simpler object to study which would yet
be able to give insight into the structures which had been
observed. Several such ``simplified'' objects have 
been studied~\cite{xxx}
 which consist of more or less forcibly modifying parts of
the integrals for the ${\hat \chi}^{(n)}$. However, there is one
``simplified''  model which commands interest in its own right. 
This is the ``diagonal susceptibility'' which is defined~\cite{mccoy3} by
 restricting the sum in (\ref{chidef}) to the correlation of spins on the
   diagonal
\begin{equation}
k_BT \cdot \chi_d\, =\,\, \sum_{N=-\infty}^{\infty}
\{\langle\sigma_{0,0}\sigma_{N,N}\rangle-M^2(0)\}.
\label{diag}
\end{equation}

In statistical language this diagonal susceptibility  is 
the susceptibility for a magnetic field interacting only with 
the spins on one diagonal. In magnetic language this is 
the $p=0$ value of the groundstate structure function
\begin{equation}
S^x(p)\, =\, \,
 \sum_{j=-\infty}^{\infty}e^{ipj}\{\langle\sigma^x_0\sigma_j^x\rangle-M_x^2\}
\end{equation}
of the transverse Ising model 
\begin{equation}
H_{TI}\, =\,\, 
 \sum_{j=-\infty}^{\infty}\{\sigma^x_{j}\sigma^x_{j+1}+H^z\sigma^z_j\}
\end{equation}

These interpretations give the diagonal susceptibility a physical
interpretation which the other ``simplified'' models do not 
have~\cite{xxx}.
Furthermore  much more analytic information is available for the
diagonal Ising correlations than for correlations off the diagonal. Firstly
it is known from the work of Jimbo and Miwa~\cite{jm} that the  diagonal
correlations are characterized by the solutions of a particular sigma
form of Painlev{\'e} VI equation and secondly the integral representation of
the diagonal form factors (\ref{dffm}) and (\ref{dffp}) is more tractable
than the representation  of the general off diagonal correlations.

The diagonal form factors have been extensively studied
 in~\cite{mccoy1} by means of processing the differential equations
obtained from long series expansions by use of Maple. Diagonal form factors
 $f^{(n)}_{N,N}$ for $n$ as large as 9 and $N$ as large as 4 have been
studied and many examples are given in~\cite{mccoy1} where they have
all been reduced to expressions in the elliptic integrals $E$ and $K$.
A few such examples are as follows:

For $n=\, 1$ (when the hypergeometric function of (\ref{f1nn}) is reduced
to the basis of $E$ and $K$ by use of the contiguous relations)

\begin{eqnarray}
&&f^{(1)}_{0,0}\,=\, \,(2/\pi)\cdot K \label{100}\\
&&t^{1/2}\, f^{(1)}_{1,1}\,=\,\, (2/\pi)\cdot \{K-E\}\\
&&3\,t\,f^{(1)}_{2,2}\,=\,\,(2/\pi) \cdot \{(t+2\,)K\, \, -2(t+1)\,E\}\\
&&15\,t^{3/2}\,f^{(1)}_{3,3}\,=\,\,(2/\pi) \cdot\{(4t^2+3t+8)\,K\,-(8t^2+7t+8)\,E\}\\
&&105\,t^2\,f^{(1)}_{4,4}\,=\,\,(2/\pi) \cdot \{(24t^3+17t^2+16t+48)K\nonumber\\
&&~~~~~~~~~~~-(48t^3+40t^2+40t+48)\,E\};
\end{eqnarray}
for $n=2$
\begin{eqnarray}
&&2\,f^{(2)}_{0,0}\,=\,\,(2/\pi)^2 \cdot K\, (K-E)\label{200}\\
&&2\,f^{(2)}_{1,1}\,=\,\,1\,\, -(2/\pi)^2 \cdot K \cdot \{(t-2)\, K \, +3E\}\\
&&6\,t\,f^{(2)}_{2,2}\,=\,6t\nonumber\\
&&-(2/\pi)^2\cdot \{6t^2-11t+2)\,K^2\,+(15t-4)\,KE\,+2(t+1)\,E^2\}\\
&&90\,t^2\,f^{(2)}_{3,3}\,=\,\, 135t^2-(2/\pi)^2\cdot \{(137t^3-242t^2+52t+8)\,K^2\\
&&-(8t^3-319t^2+122t+16)\,KE\,+4(t+1)(2t^2+13t+2)E^2\}\\
&&3150\,t^3\,f^{(2)}_{4,4}\,=\,\,6300\, t^2\nonumber\\
&&-(2/\pi)^2\cdot \{(32t^5+6440t^4-1119t^3+2552t^2+464t+128)\, K^2\nonumber\\
&&-(128t^5+576t^4-14519t^3+548t^2+1056t+256)\, KE\nonumber\\
&&+(1+t)(16t^4+58t^3+333t^2+58t+16)\,E^2\};
\end{eqnarray}
for $n=3$
\begin{eqnarray}
&&6\,f^{(3)}_{0,0}\,=\,\,(2/\pi)\cdot \,K\, \, -(2/\pi)^3\cdot \,K^2 \cdot \{(t-2)\,K\,+3E\}
\label{300}\\
&&6\,t^{1/2}\,f^{(3)}_{1,1}\,=\,\,4 \, (2/\pi)
\cdot (K-E) -(2/\pi)^3 \cdot K \cdot \{(2t-3)\,K^2\,+6KE\,-3E^2\}\nonumber\\
&&\\
&&18\,t\, f^{(3)}_{2,2}\,=\,\,7 \,\,  (2/\pi)\cdot \{(t+2)\, K \, -2\, (t+1) \, E\}\\
&&-(2/\pi)^3\cdot \{3(t^2-2) \, K^3 \, -3 \, (2t^2-11t+2)\,K^2E\nonumber\\
&&-36 \, (t-1)\,KE^2\,-24E^3\}\\
&&270 \, t^{5/2} \, f^{(3)}_{3,3}\,=\,\,30\,\,  (2/\pi)\{(4t^2+3t+8)\,K\,
-(8t^2+7t+8)\,t\,E\}
\nonumber\\
&&-(2/\pi)^3 \cdot \{(72t^4-158t^3+189t^2-156t+8)\,K^3\nonumber\\
&&-6(24t^4-108t^3+29t^2-6t+4)\,K^2E\nonumber\\
&&-3\, (232t^3\, -111t^2-180\,t\, -8)\,KE^2\nonumber\\
&&-4 \, (t+1)(2t^2+103t+2)\,t\,E^3\};
\end{eqnarray}
for $n=4$
\begin{eqnarray}
&&24\,f^{(4)}_{0,0}\,=\,\,4 \, (2/\pi)^2 \cdot K\,(K-E)\nonumber\\
&&-(2/\pi)^4 \cdot K^2\{(2t-3)\,K^2\,+6KE\,-3E^2\}\\
&&24\,f^{(4)}_{1,1}\,=\,\,9\, \, -(2/\pi)^2 \cdot 10\, K\{(t-2)\,K\,+3E\}\nonumber\\
&&+(2/\pi)^4 \cdot K^2\{(t^2-6t+6)\, K^2\, +10\, (t-2)\, KE\, +15E^2\}\\
&&72\,t\,f^{(4)}_{2,2}\,=\,\,72t\nonumber\\
&&-(2/\pi)^2 \cdot 16\cdot \{(6t^2-11t+2)\,K^2\,+(15t-4)\,KE\,+2(t+1)\,E^2\}
\nonumber\\
&&+(2/\pi)^4 \cdot \{24t^3-98t^2+113t-36)\,K^4\,+2\, (74t^2-157t+66)\,K^3E\nonumber\\
&&+3 \, (71t-60)\, K^2E^2\,+12\, (t+9)\,KE^3\,-24E^4\}.
\end{eqnarray}

These examples are sufficient to illustrate the following phenomena
which hold for all examples considered in ~\cite{mccoy1} and which are
certainly true in general:

\begin{eqnarray}
&&f_{N,N}^{(2n)}\,\,=\,\,\,\sum_{j=0}^{n} \, c^-_{j;n}\cdot g^{(2j)}_{N,N}(t)\label{sumeven}\\
&&f_{N,N}^{(2n+1)}\,\,=\,\,\,\sum_{j=0}^n \, c^+_{j;n}\cdot g^{(2j+1)}_{N,N}(t)\label{sumodd}
\end{eqnarray}
where  $c^{\pm}_{j;n}$ are constants independent of $t$ and
$g^{(j)}_{N,N}(t)$ for even $j$ are of the form
\begin{eqnarray}
&&g^{(2n)}_{0,0}(t)\,\,=\,\,\,\sum_{j=0}^n\,  P^-_{j,n;0}(t) \cdot K^{2n-j}\,E^{j} \\
&&g^{(2n)}_{1,1}(t)\,\,=\,\,\,\sum_{j=0}^n \, P^-_{j,n;1}(t) \cdot K^{2n-j}\,E^{j}\\
&&g^{(2n)}_{N,N}(t)\,\,=\,\,\, t^{-N+1}\sum_{j=0}^{2n} \, P^-_{j,n;N}(t) \cdot K^{2n-j}\,E^j~~~~
{\rm for}~~~~N\geq 2
\end{eqnarray}
and for odd $j$
\begin{eqnarray}
&&g_{0,0}^{(2n+1)}(t)\,=\,\,\sum_{j=0}^n \, P^+_{j,n;0}(t) \cdot  K^{2n+1-j}\,E^j\\
&&g^{(2n+1)}_{1,1}(t)\,=\,\,\,t^{-1/2}\cdot \sum_{j=0}^{n+1} \, P^+_{j,n;1}(t) \cdot K^{2n+1-j}\,E^j\\
&&g^{(2n+1)}_{N,N}(t)\,=\,\,t^{-N/2}\cdot \sum_{j=0}^{2n+1} \, P^+_{j,n;N}(t) \cdot K^{2n+1-j}\,E^j~~~~
{\rm for}~~~~N\geq 2
\end{eqnarray}
where $P^{\pm}_{j,n;m}(t)$ are polynomials.

The decompositions (\ref{sumeven}) and (\ref{sumodd}) represent  a
direct sum decomposition of the form factors ~\cite{mccoy3}.
The functions $g^{(j)}_{N,N}$ individually are annihilated by 
Fuchsian operators which are equivalent
to the $j+1$ symmetric power of the  second order operator
assosciated with the complete
elliptic integral $E$ (or equivalently $K$).


We also observe the relation between $f_{1,1}^{(n)}(t)$ and
$f_{0,0}^{(n+1)}(t)$
\begin{eqnarray}
&&(2/\pi)\cdot  K \cdot f_{1,1}^{(2n)}(t) 
\, =\,\, \, (2n+1) \cdot  f^{(2n+1)}_{0,0}(t) \label{rel1}\\
&&(2/\pi)\cdot  t^{1/2} \cdot K \cdot f_{1,1}^{(2n+1)}(t) 
\, = \, \,  2\,\, (n+1) \cdot f_{0,0}^{(2n+2)}(t) \label{rel2}
\end{eqnarray}

\section{Nome  $q$-representation versus modulus $k$-representation }

We will need the following identities which  relate functions of the
nome $q=\, e^{i\pi \tau}$ where $\tau=\, iK(k')/K(k)$
with functions of the modulus $k$
\begin{equation}
k \, =\, \, \frac{\theta_2^2}{\theta_3^2}, \quad \quad \quad 
k' \, = \, (1-k^2)^{1/2}\, =\,\, \, \frac{\theta_4^2}{\theta_3^2},
\label{id2}
\end{equation}
\begin{equation}
\frac{2}{\pi}\, K\, =\,\, \theta_3^2, \quad \quad  \hbox{and} \quad \quad 
\frac{dq}{dk}\,=\,\, \frac{\pi^2}{2}\frac{q}{kk'^2K^2}
\label{id4}
\end{equation}
which we will use as
\begin{equation}
q \, \frac{d}{dq}\,=\,\, \, \frac{2}{\pi^2}\,k\, k'^2\cdot K^2 \cdot \frac{d}{dk}.
\label{id5}
\end{equation}
We will also use
\begin{eqnarray}
\frac{dK}{dk}\, =\,\, \,  \frac{E-k'^2K}{kk'^2} \label{dK}, \qquad  \quad  
\frac{dE}{dk}\, =\,\,  \, \frac{E-K}{k} \label{dE}
\end{eqnarray}

\subsection{$f^{(2n)}_{0,0}$}

We first write (\ref{f00}) for $j=2n$ using (\ref{id2}) as

\begin{equation}
f^{(2n)}_{0,0}\, = \,\,\, \frac{1}{\theta_4}
\cdot \Phi_{0}\Bigl( \frac{2^{2n}\, z^{2n}\cdot (1-z^2)}{(1+z^2)^{2n+1}} \Bigr)
\end{equation}
Thus, by use of the elementary  expansion
\begin{equation}
\frac{2^{2n}\cdot z^{2n}\cdot (1-z^2)}{(1+z^2)^{2n+1}}
\, =\,\,\,  2\, \,  \frac{(-1)^n}{(2n)!}\sum_{j=0}^{\infty}(-1)^j\, z^{2j}
\prod_{m=0}^{n-1}\, 4\, [j^2-m^2]
\end{equation}
and the definition (\ref{phidef}) of the operator $\Phi_0$, 
we find that in terms of the nome $q$
\begin{eqnarray}
&&f^{(2n)}_{0,0}\, 
=\, \, 2\,\,  \frac{(-1)^n\, 4^n}{\theta_4(2n)!} \cdot 
\sum_{j=0}^{\infty}(-1)^j\, q^{j^2}\prod_{m=0}^{n-1}[j^2-m^2]\nonumber\\
&&=\, \, 2\frac{(-1)^n\, 4^n}{\theta_4(2n)!} \cdot 
\sum_{j=0}^{\infty}(-1)^j
\prod_{m=0}^{n-1}[q\, \frac{d}{dq}-m^2]q^{j^2}\nonumber\\
&&=\, \frac{(-1)^{n}\, 4^n}{\theta_4(2n)!}
\cdot \prod_{m=1}^{n-1}[q\, \frac{d}{dq}\, -m^2]\cdot q\frac{d}{dq}\theta_4.
\label{f2n00}
\end{eqnarray}

To convert this to an expression in terms of the modulus $k$ we first use 
(\ref{id2}) 
 to write
\begin{equation}
\theta_4^2\,\, =\, \, \,\frac{2}{\pi} \cdot  k' \cdot K
\label{id6}
\end{equation}
and thus using (\ref{id4})
\begin{equation}
q\frac{d}{dq}\theta_4^2\,\, =\, \, \,\frac{2}{\pi^2} \cdot  k\, k'^2\cdot  K^2
\cdot \frac{d}{dk}\left(\frac{2}{\pi}k'K\right)
\end{equation}
which using (\ref{dK}) reduces to 
\begin{equation}
q \, \frac{d }{dq}\theta_4^2\,\, 
=\,\, \, \frac{2}{\pi^2} \cdot  k' \cdot  K^2 \cdot \frac{2}{\pi}\{E-K\}.
\end{equation}
Using (\ref{id6}) on the right  hand side we find
\begin{equation}
2\, \theta_4\cdot  q\frac{d}{dq}\theta_4
\, =\,\,  \frac{2}{\pi^2} \cdot  \theta_4^2\cdot  K \cdot  \{E-K\}
\end{equation}
and thus
\begin{equation}
\frac{1}{\theta_4}\cdot  q\, \frac{d}{dq}\theta_4 \, 
=\, \, \frac{1}{\pi^2} \cdot  K \cdot \{E-K\}.
\label{qdqt}
\end {equation}

To evaluate $f^{(2)}_{0,0}$ we use (\ref{qdqt}) in (\ref{f2n00}) with
$n=1$ to obtain
\begin{equation}
f^{(2)}_{0,0}\, =\,\,  \frac{2}{\pi^2}\cdot  K \cdot \{K-E\}
\end{equation}
which is in agreement with  (\ref{200}). For arbitrary $n$ the form factor
$f^{(2n)}_{0,0}$ is obtained from (\ref{f2n00}) by repeated use of
(\ref{qdqt}) and (\ref{dK}).
 
\subsection{$f^{(2n+1)}_{0,0}$}

To study $f^{(2n+1)}_{0,0}$ we first use 
(\ref{id2})  to write (\ref{f00}) as
\begin{equation}
f^{(2n+1)}_{0,0}\, =\,\,\, 
\frac{\theta_3}{\theta_2\theta_4} \cdot 
\Phi_0\Bigl( \frac{2^{2n+1}\, z^{2n+1}\cdot (1-z^2)}
{(1+z^2)^{2n+2}}\Bigr) 
\end{equation}
and then, using the elementary expansion
\begin{eqnarray}
&&\frac{2^{2n+1} \cdot z^{2n+1} \cdot (1-z^2)}{(1+z^2)^{2n+2}} \\
&&=\,\, 2 \,\, \frac{(-1)^n}{(2n+1)!} \cdot 
\sum_{j=0}^{\infty} \, (2j+1) \cdot  (-1)^j \cdot  z^{2j+1}
 \cdot  \prod_{m=0}^{n-1}[(2j+1)^2-(2m+1)^2]\nonumber
\end{eqnarray}
and the definition (\ref{phidef}) of the operator $\Phi_0$ we find
\begin{eqnarray}
&&f^{(2n+1)}_{0,0}\nonumber   \\
&&=\frac{\theta_3}{\theta_2\theta_4}\frac{2(-1)^n}{(2n+1)!} \cdot 
\sum_{j=0}^{\infty}(2j+1) \cdot (-1)^j \cdot q^{(2j+1)^2/4}\cdot 
\prod_{m=0}^{n-1}[(2j+1)^2-(2m+1)^2]\nonumber\\
&&~~~~~=\frac{\theta_3}{\theta_2\theta_4}\frac{2\, \, (-1)^n}{(2n+1)!}
\cdot \prod_{m=0}^{n-1}[4 \, q\frac{d}{dq}-(2m+1)^2]\, 
\sum_{j=0}^{\infty}(2j+1)(-1)^j q^{(2j+1)^2/4}.  
\end{eqnarray}
Thus, if we write
\begin{equation}
2\, \sum_{j=0}^{\infty} \, (2j+1)\cdot  (-1)^j \cdot  q^{(2j+1)^2/4} \,\,  = \, \, 
\frac{\partial}{\partial
  u}\theta_1(u,q)|_{u=0} \,\,  =\, \, \theta_2\theta_3\theta_4
\end{equation}
where in the last line we have used a well known identity,
we find the result
\begin{equation}
f^{(2n+1)}_{0,0}\, =\,\,  \frac{\theta_3}{\theta_2\theta_4}\frac{(-1)^n}{(2n+1)!}
\cdot \prod_{m=0}^{n-1} \, [4 \, q\frac{d}{dq} 
\, -(2m+1)^2] \, \, \theta_2\theta_3\theta_4.
\label{f2np100}
\end{equation}

We may now use (\ref{id2})--(\ref{dE}) to reduce (\ref{f2np100}) from a
function of $q$ to a function of $k$. 

For $n=\, 0$ we  use (\ref{id4}) to find
\begin{equation}
f^{(1)}_{0,0} \, =\, \, \theta_3^2 \, = \, \, {{2} \over {\pi}} \cdot  K
\end{equation}
which agrees with (\ref{100}).

For $n \geq \, 3$ we need an expression analogous to (\ref{qdqt}) for 
the product $\theta_2\theta_3\theta_4$. From (\ref{id2}), (\ref{id4})
\begin{eqnarray}
\theta_2^2\theta_3^2\theta_4^2 \,= \, k \,k' \cdot  (2 \,K/\pi)^3
\end{eqnarray}
and thus
\begin{eqnarray}
&&q\frac{d}{dq}\theta_2^2\theta_3^3\theta_4^2 \, 
=\, \, 2 \, \theta_2\theta_3\theta_4\cdot  q\frac{d}{dq}\theta_2\theta_3\theta_4\,
=\, \frac{2}{\pi^2}  \cdot k \,k'^2 \,K^2 \cdot \frac{d}{dk}\{k \,k'(2\,K/\pi)^3\}
\nonumber\\
&&=\,\frac{2}{\pi^2} \cdot k \,k'  \cdot (2K/\pi)^3 \cdot K \cdot \{(k^2-2)\,K\,+3E\}
\nonumber\\
&&=\,\frac{2}{\pi^2} \cdot \theta^2_2\, \theta^2_3\,
\theta^2_4 \cdot K \cdot \{(k^2-2)\,K\,+3E\}.
\end{eqnarray}
Therefore we obtain 
\begin{equation}
q\frac{d}{dq}\theta_2\theta_3\theta_4
\, =\,\,\frac{1}{\pi^2}\theta_2\theta_3\theta_4 \cdot K\cdot \{(k^2-2)\,K\,+3E\}
\end{equation}
which when used in (\ref{f2np100}) with $n=1$ gives
\begin{equation}
f^{(3)}_{0,0}\,=\,\, 
 \frac{1}{3!}\cdot \{(2/\pi)\cdot K\,\, 
-(2/\pi)^3 \, K\cdot [(k^2-2)\, K \, +3E]\}
\end{equation}
which is in agreement with (\ref{300}).

\subsection{$f^{(n)}_{1,1}$}

The equalities (\ref{rel1}) and (\ref{rel2}) which express
$f^{(n)}_{1,1}$ in terms of $f^{(m)}_{0,0}$ follow immediately from
  (\ref{f00}) and (\ref{f11}) by use of (\ref{id2}) and (\ref{id4}). 

\section{The $\lambda$ generalized correlations}

For $N=0,1$ the diagonal $\lambda$ generalized
correlations defined by (\ref{lffm}) and (\ref{lffp}) may be obtained
by using the expressions for $(1-t)^{1/4}f^{(n)}_{N,N}$ in terms of the
operator $\Phi_{0}$ as given by (\ref{f00}) and (\ref{f11}). In this
form the sums over $n$ are easily done as geometric series and the linear
operator $\Phi_0$ is then used to convert the series in $z$ to series
in the nome $q$ which can then be expressed in terms of $\theta$
functions as was done in the previous section. Then, setting
\begin{equation}
\lambda\,=\,\, \cos u
\end{equation}
we obtain the following results
\begin{eqnarray}
&&C_{-}(0,0;\lambda)\,=\, \,\frac{\theta_3(u;q)}{\theta_{3}(0;q)}\label{gen0m}\\
&&C_{+}(0,0;\lambda)\,=\,\,\frac{\theta_2(u;q)}{\theta_2(0;q)}\label{gen0p}\\
&&C_{-}(1,1;\lambda)\,=\,\,\frac{-\theta'_2(u;q)}
{\sin (u)\,\theta_2(0;q)\,\theta_3(0;q)^2}\label{gen1m}\\
&&C_{+}(1,1;\lambda)\,=\,\, \frac{-\theta'_3(u;q)}
{\sin (u)\, \theta^2_2(0;q)\,\theta_3(0;q)}\label{gen1p}
\end{eqnarray}
where prime indicates the derivative with respect to $u$.
The result (\ref{gen0m}) was first reported in~\cite{mccoy1}. 
The results (\ref{gen0p})--(\ref{gen1p}) have recently 
been given in \cite{tony}.
For $u$ a rational multiple of $\pi$ these diagonal generalized
correlations  reduce to algebraic functions of the modulus
$k$. Several examples for $C_{-}(0,0;\lambda),\, \, C_{-}(1,1;\lambda)$ and 
$C_{-}(2,2;\lambda)$
are given in~\cite{mccoy1}.

\section{Diagonal susceptibility}

We may now explicitly obtain~\cite{mccoy3} the diagonal susceptibility
by using the form factor expansion (\ref{ffm})--(\ref{dffp}) in the
definition (\ref{diag}) and
evaluate the sum on $N$ as a geometric series. We obtain for $ \, T < T_c$
\begin{eqnarray}
kT \cdot \chi_{d-}(t)\,\,  =\,\, \,  \, (1-t)^{1/4} \cdot 
\sum_{n=1}^{\infty} \, {\tilde \chi}_d^{(2n)}(t)
\label{defchidm}
\end{eqnarray}
with 
\begin{eqnarray}
&&{\tilde \chi}^{(2n)}_{d}(t)\,\, = \,\,\,  \, \,
 {{  t^{n^2}} \over {  (n!)^2 }} \, 
{{1 } \over {\pi^{2n} }} \cdot  
\int_0^1 \cdots  \,\int_0^1\prod_{k=1}^{2n}\,  dx_k  
\cdot   {1\, +t^n\, x_1\cdots x_{2n}\over
  1\,-t^n\, x_1 \cdots x_{2n}}\nonumber \\ 
&&\quad \quad \quad \quad \times
\prod_{j=1}^n\left({x_{2j-1}(1-x_{2j})(1-tx_{2j})\over 
x_{2j}(1-x_{2j-1})(1\, -t\, x_{2j-1})}\right)^{1/2}\nonumber\\
&&\quad \quad \quad \quad \times \prod_{1 \leq j \leq n}
\prod_{1 \leq k \leq n}(1\, -t\, x_{2j-1}\, x_{2k})^{-2} \nonumber\\
&&\quad\quad \quad \quad  \times
\prod_{1 \leq j<k\leq n}(x_{2j-1}-x_{2k-1})^2\, (x_{2j}-x_{2k})^2
\label{chidm}
\end{eqnarray}
and for $T>T_c$
\begin{eqnarray}
\label{defchidp}
kT \cdot \chi_{d+}(t)\, =\,\,\,  (1-t)^{1/4} \cdot
 \sum_{n=0}^{\infty}\, {\tilde \chi}_d^{(2n+1)}(t)
\end{eqnarray}
with 
\begin{eqnarray}
&&{\tilde \chi}^{(2n+1)}_{d}(t)\,\, = \,\,\,\,\,\,{t^{n(n+1))}
 \over \pi^{2n+1} n! \, (n+1)! } \cdot 
\int_0^1 \cdots \int_0^1 \, \, \prod_{k=1}^{2n+1}  dx_k\nonumber\\
&&\quad \quad \quad \quad  \times {1\, +t^{n+1/2}\,x_1\cdots x_{2n+1}\over
 1\, -t^{n+1/2}\, x_1\cdots x_{2n+1}} \cdot 
 \prod_{j=1}^{n}\, \Bigl((1-x_{2j})(1\,-t\,x_{2j})\cdot
 x_{2j}\Bigr)^{1/2} \nonumber \\
&&\quad \quad \quad \quad \times\prod_{j=1}^{n+1} \, 
\Bigl((1\,  -x_{2j-1})(1\,-t\, x_{2j-1}) \cdot x_{2j-1}\Bigr)^{-1/2} 
\nonumber\\
\label{chidp}
&&\quad \quad\quad  \quad  \times
\prod_{1\leq j\leq n+1}\prod_{1\leq k \leq n}\, 
(1\, -t\, x_{2j-1}\, x_{2k})^{-2} \\
&&\quad \quad  \quad \quad \times\prod_{1 \leq j<k\leq n+1}(x_{2j-1} -x_{2k-1})^2
\prod_{1\leq j<k\leq n}(x_{2j}-x_{2k})^2. \nonumber
\end{eqnarray} 

This diagonal susceptibility has been extensively studied in~\cite{mccoy3}. 

The integrals for ${\tilde \chi}^{(1)}_d(t)$ and ${\tilde \chi}^{(2)}_d(t)$
are explicitly evaluated as
\begin{eqnarray}
\label{chi12}
{\tilde \chi}^{(1)}_d(t)\,=\,\,\,\, {1 \over 1\,\, -t^{1/2}}
\end{eqnarray}
and
\begin{eqnarray}
\label{chi24}
{\tilde \chi}^{(2)}_d(t)
\,\,=\,\,\, {1\over 8\pi i}\oint dz_1 
\,{t \over (1\,-t^{1/2}z_1)(z_1\,-t^{1/2})}
\,\,=\,\,\,\,{t \over 4\,(1-t)}. 
\end{eqnarray}

Fuchsian equations have been obtained for  ${\tilde \chi}_d^{(3)}(t)$,
${\tilde \chi}_d^{(4)}(t)$  and ${\tilde \chi}_d^{(5)}(t)$. 

From these equations we find that ${\tilde \chi}_d^{(3)}(t)$  has a
direct sum decomposition into the sum of three terms. One term is just 
${\tilde \chi}^{(1)}_d(t)$ as given by (\ref{chi12}); the second is 
\begin{equation}
\frac{1}{k-1} \cdot {{2} \over {\pi}} \, K
\, \, \,+\frac{1}{(k-1)^2} \cdot {{2} \over {\pi}}\, E
\end{equation} 
and the three solutions to the differential equation for the third term 
are two Meijer G functions and 
\begin{eqnarray}
&&\frac{(1+2k)(k+2)}{(1-k)(1+k+k^2)} \cdot \{F(1/6,1/3;1;Q)^2
\nonumber\\
&& ~~~~~~~~~~~+\frac{2Q}{9} \cdot F(1/6, 1/3;1;Q)F(7/6,4/3;2;Q)\}
\end{eqnarray}
where
\begin{equation}
Q \, = \, \,
\frac{27}{4}\frac{(1+k)^2k^2}{(k^2+k+1)^2}
\end{equation}

Furthermore the ${\tilde \chi}_d^{(n)}(t)$ have singularities on
$|t|=1$ which are the analog of the Nickel singularities for the
bulk susceptibility. For $T<T_c$ the singularities in 
${\tilde  \chi}^{(2n)}(t)$
are at $t^n=1$ and are of the form $\epsilon^{2n^2-1}\ln \epsilon$ 
and for $T>T_c$ the singularities in ${\tilde \chi}^{(2n+1)}$
 are at $\, t^{n+1/2}=1$ and are of the form
  $\, \epsilon^{(n+1)^2-1/2}.$

\section{Natural boundary}

The most intriguing feature in both the bulk and the diagonal
susceptibility are the singularities on the unit circle of the modular
variable $k=1$. As $n$ increases the number of these singularities
increases and becomes dense as $n\, \rightarrow\, \infty$. Therefore,
unless a massive cancellation occurs the susceptibility will have a
natural boundary on the circle $|k|=1.$ Recently further arguments in
favor of such a natural boundary were given in \cite{maillard4}.
In terms of the nome $q$ the
circle $|k|=1$ corresponds to the curve in Fig. 1.

\begin{figure}[here]
\center
\epsfig{file=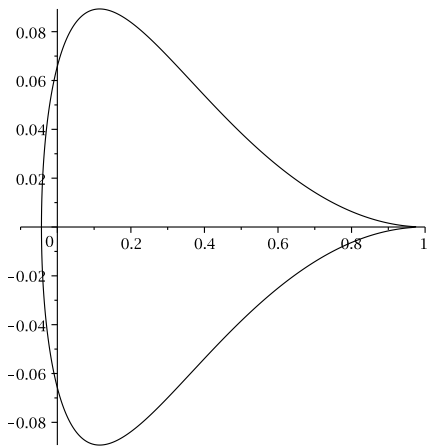,width=6cm,height=6cm}
\caption{The curve in the plane of the nome $q$ of the unit circle $|k|=1$}
\label{fig1}
\end{figure}

\section{Conclusion}

We have seen that since 1959 a great deal of progress has been made in
understanding the susceptibility of the Ising model and that the
analytic structure is vastly more complicated than was envisaged 50
years ago in~\cite{fisher}. In particular the existence of a natural
boundary is a completely new phenomenon which has never before
appeared in the study of critical behavior. The connections with
elliptic modular functions are profound and extensive and much of the
structure still remains to be discovered. It is quite remarkable  that
in 50 years the problem has not been solved.
 
\section*{Acknowledgments}

One of us, BMM, wishes to thank the organizers  for the opportunity to
participate in ``New trends in
quantum integrable systems'' in honor of the 60th birthday of Tetsuji
Miwa. JMM thanks the Simons Center for Geometry and Physics
for support which helped in the finalizing of the manuscript.
 BMM also wishes to thank Prof. Miwa for 35 years worth in inspiration.


\begin{thebibliography}{99}

\bibitem{ons1} L. Onsager, Crystal statistics,I. A two-dimensional model
  with an order disorder transition,  Phys. Rev. 65 (1944) 117--149.

\bibitem{ons2} L. Onsager, discussion, Nuovo Cimento 6 suppl. (1949) 261.

\bibitem{yang} C.N. Yang, The spontaneous magnetization of the two
  dimensional Ising model, Phys. Rev. 85 (1952) 808--816.

\bibitem{fisher} M.E. Fisher, The susceptibility of the plane Ising
  model, Physics 25 (1959) 521-524.

\bibitem{ko}B. Kaufmann and L. Onsager, Crystal statistics III. short
  range order in a binary Ising lattice, Phys. Rev. 76 (1949) 1244--1252. 

\bibitem{sn} I. Syozi and S. Naya, 
Symmetrical properties of two-dimensional Ising lattices, 
Prog. Theo. Phys. 24 (1960) 829--839.


\bibitem{mpw} E.W. Montroll, R.B. Potts and J.C. Ward, Correlations
  and spontaneous magnetization of the two dimensional Ising model,
  J. Math. Phys. 4 (1963) 308-322.

\bibitem{wu} T.T. Wu, Theory of Toeplitz determinants and spin
  correlations of the two dimensional Ising model, Phys. Rev. 149
  (1966) 380--401.


\bibitem{wmtb} T.T. Wu. B.M. McCoy, C.A. Tracy and E. Barouch,
  Spin-spin correlation functions for the two dimensional Ising model:
  exact theory in the scaling region, Phys. Rev. B13 (1976) 315--374.

\bibitem{pt} J. Palmer and C. Tracy, Two-dimensional Ising
  correlations: convergence of the scaling limit,  
Adv. Appl. Math. 2 (1981) 329--388.

\bibitem{yam} K. Yamada, 
On the spin-spin correlation functions in the Ising square lattice and
the zero field susceptibility,
Prog. Theo. Phys. 71 (1984) 1416-1421.

\bibitem{mccoy1} S. Boukraa, S. Hassani, J-M. Maillard, B.M. McCoy,
  W.P. Orrick and N. Zenine, Holonomy of the Ising model form factors,
  J. Phys. A40 (2007) 75-112.

\bibitem{mccoy2} I. Lyberg and B.M. McCoy, Form factor expansion of
  the row and diagonal correlation functions of the two dimensional
  Ising model, J. Phys. A 40 (2007) 3329--3346.


\bibitem{mtw} B.M. McCoy, C.A. Tracy and T.T. Wu, Painlev{\'e}
  equations of the third kind, J. Math. Phys. 18 (1977) 1058--1092.

\bibitem{bmw} E. Barouch, B.M. McCoy and T.T. Wu, Zero field
  susceptibility of the two dimensional Ising model near $T_c$,
  Phys. Rev. Letts. 31 (1973) 1409-1411.


\bibitem{ongp} W.P. Orrick, B.G. Nickel, A.J. Guttmann and
  J.H.H. Perk, The susceptibility of the square lattice Ising model:
  new developements, J. Stat. Phys. 102 (2001) 795--841.


\bibitem{ge}A.J. Guttmann and I.G. Enting, Solvability of some
  statistical mechanical systems, Phys. Rev. Letts. 76 (1996) 344--347.


\bibitem{nickel1} B.G. Nickel, On the singularity structure of the 2D Ising
  model, J. Phys. A32 (1999) 3889--3906.

\bibitem{nickel2} B.G. Nickel, Addendum to 
``On the singularity structure of the 2D Ising model'', J. Phys. A 33
  (2000) 1693--1711.


\bibitem{maillard1} N. Zenine, S. Boukraa, S. Hassani and J-M.
  Maillard, The Fuchsian differential equation of the square Ising
  $\chi^{(3)}$ susceptibility, J. Phys. A 37 (2004) 9651--9668.

\bibitem{maillard2} N. Zenine, S. Boukraa, S. Hassani and J-M.
  Maillard, Square lattice Ising model susceptibility: series expansion
  method and differential equation for $\chi^{(3)}$, J. Phys. A 38
  (2005) 1975--1899.

\bibitem{maillard3} N. Zenine, S. Boukraa, S. Hassani and J-M.
  Maillard, Ising model susceptibility: The Fuchsian equation for
  $\chi^{(4)}$ and its factorization properties, J. Phys. A 38
  (2005) 4149--4173.

\bibitem{maillard4} S. Boukraa, A.J. Guttmann, S. Hassani, I. Jensen,
  J-M. Maillard, B. Nickel and N. Zenine, Experimental mathematics on
  the magnetic susceptibility of the square lattice Ising model,
  J. Phys. A 41 (2008) 455202.

\bibitem{maillard5} A. Bostan, S. Boukraa, A.J. Guttmann, S. Hassani,
  I. Jensen, J-M. Maillard and N. Zenine, High order Fuchsian
  equations for the square lattice Ising model: ${\tilde \chi}^{(5)}$,
J. Phys. A 42 (2009) 275209.

\bibitem{maillard7} S. Boukraa, S. Hassani, I. Jensen, J-M. Maillard, 
N.J. Zenine,  High order Fuchsian equations for the square lattice 
Ising model: $\, \chi^{(6)}$, 
 in press J. Phys.A: Math. Theor. (2010), arXiv:0912.4968v1 [math-ph]. 


\bibitem{maillard6} A. Bostan, S. Boukraa, S. Hassani, J-M. Maillard,
  J-A. Weil and N. Zenine, Global nilpotent differential operators and
  the square Ising model, J. Phys. A 40 (2007) 2583--2614.


\bibitem{xxx} S. Boukraa, S. Hassani, J-M. Maillard, and N. Zenine,
  Landau singularities and singularities of holonomic integrals of the
  Ising class, J. Phys. A 40 (2007) 2583-2614


\bibitem{mccoy3} S. Boukraa, S. Hassani, J-M. Maillard, B.M. McCoy, 
J-A. Weil and
  N. Zenine, The diagonal Ising susceptibility, J. Phys. A 40 (2007)
  8219--8236.


\bibitem{jm}
M. Jimbo and T. Miwa, Studies on holonomic quantum fields XVII,
Proc. Jpn. Acad. 56A (1980) 405: 57A (1981) 347.

\bibitem{tony} V.V. Mangazeev and A.J. Guttmann, Form factor expansions
  in the 2D Ising model and Painlev{\'e} VI, ArXiv.1002.2490.

\end{thebibliography}
\end{document}